	\def\stalk{\vrule height1.35ex width0.03em depth-0.035ex}
	\def\halfarc{{\hbox{$\stalk\kern -0.17 em _\bullet$}}}
	\def\lowstalk{\vrule height0.85ex width0.03em depth0.75ex}
	\def\arc{{\hbox{$\lowstalk\kern -0.17 em _\bullet$}}}
	\def\higharc{{\hbox{$\vert \kern-0.38 em \raise 0.13 ex
		\hbox{$\bullet$}$}}}
        
 	\newcommand{\BOX}{\hbox {$\sqcap$ \kern -1em $\sqcup$}}
	\newcommand{\qed}{\hskip 3em \hbox{\BOX} \vskip 2ex}
	\newcommand\Hom{{\rm Hom}}
	\newcommand\C{{\bf C}}
	\def\section#1{\vskip3em{\centerline {\bf#1}}\vskip3em}
	\newcommand\R{{\bf R}}
	\newcommand{\et}{\hspace{-0.08in}{\bf .}\hspace{0.1in}}
	\newcommand{\be}{\begin{equation}}
        \newcommand{\ee}{\end{equation}}
        \newcommand{\ba}{\begin{eqnarray}}
        
	\newcommand{\om}{\omega}

        \newcommand{\tensor}{\otimes}
        \newcommand{\maps}{\colon}
        \newcommand{\ea}{\end{eqnarray}}
        
      	\renewcommand{\L}{{\cal L}}
       
	\newcommand{\A}{{\cal A}}
	\newcommand{\G}{{\cal G}}

	\renewcommand{\H}{{\bf H}}
	\renewcommand{\a}{{\bf A}}
	\newcommand{\iso}{\cong}
	\newcommand{\tr}{{\rm tr}}
	
	\newcommand{\Diff}{{\rm Diff}}

	\newtheorem{theorem}{Theorem}
	\newtheorem{proposition}{Proposition}
	\newtheorem{lemma}{Lemma}
	\newtheorem{corollary}{Corollary}

	\documentstyle[12pt]{article}
\begin{document}

	\begin{center}
	{\bf Diffeomorphism-invariant Generalized Measures\\
        On the Space of Connections Modulo Gauge Transformations\\}
	\vspace{0.5cm}
	{\em John C. Baez\\}
	\vspace{0.3cm}
	{\small Department of Mathematics \\
	University of California\\
        Riverside CA 92521\\}
	\vspace{0.3cm}
	{\small May 8, 1993\\}
	\vspace{0.3cm}
   	{\small \it to appear in the proceedings of the Conference on Quantum
	Topology, \\
	Manhattan, Kansas, March 24-28, 1993\\}
	\vspace{0.5cm}
	\end{center}

\begin{abstract}
The notion of a measure on the space of connections modulo gauge
transformations that is invariant under diffeomorphisms of the base
manifold is important in a variety of contexts in mathematical physics
and topology.  At the formal level, an example of such a measure is
given by the Chern-Simons path integral.  Certain measures of this sort
also play the role of states in quantum gravity in Ashtekar's formalism.
These measures define link invariants, or more generally multiloop
invariants; as noted by Witten, the Chern-Simons path integral gives
rise to the Jones polynomial, while in quantum gravity this observation
is the basis of the loop representation due to Rovelli and Smolin.
Here we review recent work on making these ideas mathematically
rigorous, and give a rigorous construction of diffeomorphism-invariant
measures on the space of connections modulo gauge transformations
generalizing the recent work of Ashtekar and Lewandowski.  This
construction proceeds by doing lattice gauge theory on graphs
analytically embedded in the base manifold.
\end{abstract}

\section{Introduction}

In physics, diffeomorphism-invariant ``measures'' on the space $\A/\G$
of connections modulo
gauge transformations play an important role in the quantization of
generally covariant gauge theories.   However, these ``measures'' are often
purely formal in nature; it is a challenge to find a formulation of them
that is both mathematically rigorous and sufficiently flexible.  In this
paper we
begin by reviewing of work by Ashtekar, Isham, Lewandowksi and the author
\cite{AI,AL,Baez} on holonomy C*-algebras as an approach to this
problem.  Here the heuristic notion of a ``measure'' on $\A/\G$
is replaced by the concept of a continuous linear functional on a particular
algebra of functions on $\A/\G$.   Then,
generalizing the work of Ashtekar and Lewandowski, we
construct diffeomorphism-invariant states on holonomy C*-algebras from
certain invariants of ``multiloops,'' that is, finite collections of
loops.  This construction makes it clear that to understand the relation
between diffeomorphism-invariant gauge theories
and knot theory one should treat the space
of all links as a subset of the space of multiloops and attempt to
extend link invariants to multiloop invariants.
Interestingly, the same idea has recently shown up both in the theory of
Vassiliev invariants \cite{Baez,BN,Birman,Stanford,Vas}
and in work on quantum gravity \cite{BP,BGP,Gambini,Husain,JS}.

Two very important diffeomorphism-invariant gauge theories are
Chern-Simons theory in 3 dimensions and general relativity in 4
dimensions.   In the path-integral
approach to Chern-Simons theory, we take a 3-manifold $M$ as spacetime,
consider a principal bundle $G \to P \to M$, and calculate vacuum
expectation values as integrals with respect to the ``measure''
\[     e^{iS(A)}\, {\cal D}A, \]
where ${\cal D}A$ is the purely formal
 - that is, nonexistent - Lebesgue measure on $\A/\G$,  and
\[   S(A) = {k\over 4\pi} \int_{M} \tr (A \wedge dA + {2\over 3}A \wedge A
\wedge A) \]
is the Chern-Simons action, where $k$ is an integer.   The Chern-Simons
action is preserved by the action of orientation-preserving
diffeomorphisms of $M$ and, up to integers, by the action of gauge
transformations.  Thus at a formal level, $e^{iS(A)}{\cal D}A$ is
regarded as a measure on $\A/\G$ that is
invariant under orientation-preserving diffeomorphisms of $M$.
Given a loop $\gamma \maps S^1 \to M$, let $T(\gamma,A)$ denote the
trace of the holonomy of $A$ around $\gamma$, taking the trace
in some finite-dimensional representation of $G$.   This is a gauge-invariant
function of $A$, so it can be regarded as a function $T(\gamma)$ on $\A/\G$.
Thus given a collection of loops $\{\gamma_i\}$ and an
orientation-preserving diffeomorphism $g \maps M \to M$,
we expect that
\[       \int T(\gamma_1) \cdots T(\gamma_n) e^{iS(A)}\,{\cal D}A =
  \int T(g \circ \gamma_1) \cdots T(g\circ \gamma_n) e^{iS(A)}\,{\cal D}A \]
In other words, we expect Chern-Simons theory to give
an ``ambient isotopy invariant'' of multiloops $\gamma = \{\gamma_i\}$,
and in particular of links.

Due to
the formal nature of the Chern-Simons ``measure'' it is not surprising
that there are complications.  For links, a formal calculation by Witten
\cite{Witten} using
conformal field theory indicates that the integral must be regularized
using a framing of the link.  Taking the trace in the fundamental
representation of $G = SU(2)$, the result is an ambient isotopy
invariant of framed links known as the Kauffman bracket
\cite{Kauffman}, closely related to the Jones polynomial
\cite{Jones}.  Witten's
result has been checked perturbatively by various authors
\cite{BGP,BN,Cotta,GMM}.     For
other groups $G$ and other representations, one obtains other link
invariants.   All these invariants may be constructed rigorously
using the corresponding quantum group representations \cite{RT,Tur}.
Thus we expect a close relationship between any rigorous
formulation of the Chern-Simons path integral and the representation
theory of quantum groups.

In the connection representation of quantum gravity in 4 dimensions
\cite{Ash}, we take a 3-manifold $M$ as space, rather than spacetime,
and we take $P \to M$ to be the trivial bundle with $G = SL(2,\C)$, or
in Euclidean quantum gravity, $SU(2)$.
At a formal level, states of quantum gravity are given by ``measures''
on $\A/\G$ of the form
\[        \Psi(A)\, {\cal D}A ,\]
that are invariant under the
identity component $\Diff_0(M)$ of the diffeomorphism group and also
annihilated by operators known as the Hamiltonian constraints.    (We
ignore, in this brief sketch, the important but subtle issue of the ``reality
conditions'' in this approach to quantum gravity.)  Alternatively,
following Rovelli and Smolin \cite{RS}, we can work in the loop
representation and think of the state as a function $\widehat\Psi$ of
multiloops in $M$, where $\widehat\Psi$ is the ``loop transform'' of $\Psi$:
\[      \widehat\Psi(\gamma) =  \int T(\gamma_1) \cdots T(\gamma_n)
\Psi(A)\,{\cal D}A .\]
Here $\widehat \Psi$ will automatically be an ambient isotopy invariant of
multiloops.   A very interesting problem is to describe the
Hamiltonian constraints in terms of the loop representation and find
all multiloop invariants that are annihilated by these operators
\cite{BGP,BP,Husain,Gambini}.

In fact, there is a deep relation between Chern-Simons theory and
quantum gravity, noticed by Kodama \cite{Kodama} and subsequently
explored in many papers \cite{BaezTang,BGP,Crane,Gambini,P,Smolin}.
This is that the Chern-Simons ``measure,'' in
addition to being $\Diff_0(M)$-invariant, is annihilated by the Hamiltonian
constraint for quantum gravity with cosmological constant
\[     \Lambda = {24\pi i\over k} .\]
This fact finds its deepest explanation so far in
terms of the Capovilla-Dell-Jacobson formulation of general relativity,
in which the action is closely related to the 2nd Chern class
\cite{CDP,LNU}.

Unfortunately, much of the aforementioned work, while very
interesting, is not quite
mathematics yet, because the ``measures'' in question have not been
constructed in any rigorous sense.   They are unlikely to be Borel
measures on the space $\A/\G$ with any of its standard topologies.
In order to address this issue,
Ashtekar and Isham \cite{AI} introduced a generalization of
measures on $\A/\G$, namely, continuous linear functionals on
an algebra called the holonomy C*-algebra.  This algebra
the completion in the $L^\infty$ norm of the algebra
generated by the Wilson loops $T(\gamma)$.
Before describing this algebra in Section 3, we review some general
ideas on
functional integration in Section 2.   This material is standard but
perhaps phrased in a somewhat new manner.   In Section 4 we give a
characterization of
$\Diff_0(M)$-invariant continuous linear functionals on
the holonomy C*-algebra in terms of lattice
gauge theory on graphs embedded in $M$.
Recently, Ashtekar and Lewandowsi constructed such a continuous linear
functional - in fact, a state - using a technique that depends crucially on
working with piecewise analytic loops.
In Section 5 we use the results of the previous section to construct
many $\Diff_0(M)$-invariant states on the holonomy algebras
of bundles over manifolds $M$ of any dimension.
These examples involve an interesting interplay between
the singularity theory of analytic curves in $M$ and the harmonic
analysis of measures on spaces of the form $G^d$.
In Section 6 we sketch an extension of the work in the previous
section that is applicable only in the case of 3-dimensional manifolds.
This extension, which has not been fully worked out, is very similar to
Reshetikhin and Turaev's \cite{RT,Tur}
construction of graph invariants from quantum group representations.  We
also briefly discuss the relation between regularization and
framing-dependence of link invariants.  This section may be regarded as
a program for rigorously constructing the Chern-Simons ``measure.''

\section{Generalized Measures}

In order to understand what the diffeomorphism-invariant
 ``measures'' on $\A/\G$ in physics
might really be, it is useful to
take the stance that the use of a measure is to integrate functions.
Thus, we downplay the notion of a measure as a set
function, and emphasize its role as a linear functional:
\[                 f \mapsto \int f d\mu .\]
For spaces that are not ``too big,'' there is a well-known
correspondence between measures as set functions and measures as
linear functionals.  This is the Riesz-Markov theorem: if $X$ is a
compact Hausdorff space, there is a one-to-one correspondence between
 measures on $X$ and continuous linear functionals
on $C(X)$, the algebra of continuous complex-valued functions on $X$
equipped with the $L^\infty$ norm
\[      \|f\|_\infty = \sup_{x \in X} |f(x)|  .\]
Here, and in all that follows, by {\it measure} we really mean a finite
regular Borel measure.
The Riesz-Markov theorem assigns to each measure $\mu$ the functional
\[               f \mapsto \int_X f d\mu .\]
The deep part of the theorem is that all continuous linear functionals
$\int\maps C(X)\to \C$ are of this form.

Note that $C(X)$ is a unital C*-algebra, that is, a Banach space over
$\C$ that is an algebra with multiplicative identity equipped with a
$\ast$ operation, in this case pointwise complex conjugation, satisfying
\ba            (f + g)^\ast &=& f^\ast + g^\ast \nonumber\cr
                (\lambda f)^\ast &=& \overline \lambda f^\ast \nonumber\cr
                (fg)^\ast &=& g^\ast f^\ast  \nonumber\cr
                f^{\ast \ast} &=& f   \nonumber\cr
 		\|fg\| &\le& \|f\|\|g\|  \nonumber\cr
      		\|f^\ast f\| &=& \|f\|^2.  \nonumber\ea
The Gelfand-Naimark theorem says that conversely, every
commutative unital C*-algebra is isomorphic to $C(X)$ for some compact
Hausdorff space, unique up to homeomorphism.
Taken together, the Riesz-Markov and
Gelfand-Naimark theorems allow us to treat all of measure theory on
compact Hausdorff spaces in terms of C*-algebras and continuous linear
functionals.   For example,
a positive measure on $X$ corresponds to a positive linear functional on
$C(X)$, that is, a linear map $\int \maps C(X) \to \C$ such that
\[           \int \overline f f \ge 0.  \]
(Such functionals are automatically continuous.)
Similarly, a probability measure on $X$ corresponds to a state on $C(X)$, that
is, a positive linear functional $\int$ with $\int 1  = 1$.

These results can be generalized to the case of Hausdorff spaces that
are only locally compact.
However, the infinite-dimensional spaces arising in physics
 are typically not even locally
compact.  For example, if we take $\A/\G$ to consist of smooth
connections modulo smooth gauge transformations, with the $C^\infty$
topology, it will not be locally compact.  The same holds if we work
with connections and gauge transformations lying in appropriate Sobolev
spaces.  It is certainly possible to construct measures on spaces that
are not locally compact, Wiener measure being a famous example, but
it is sometimes more simple to consider a generalization of the
notion of measure.

Let $X$ be an arbitrary Hausdorff space and let $C_b(X)$ denote the
C*-algebra of all bounded complex continuous functions on $X$.   While a
 measure would enable us to integrate all functions in
$C_b(X)$, we may well be satisfied with being able to integrate
functions in some subalgebra of $C_b(X)$, which in physics terminology
is a class of ``distinguished observables.''   Assume that ${\bf A}_0
\subseteq C_b(X)$ is a $\ast$-subalgebra, that is, a subalgebra such
that $f \in {\bf A}_0$ implies $\overline f \in {\bf A}_0$. Assume also that
there is a linear functional $\int \maps {\bf A}_0 \to \C$, and that
$\int$ is bounded, that is, for some $C > 0$
\[                |\int f| \le C \|f\|_\infty  \]
for all $f \in {\bf A}_0$.    Then we say that $\int$ is a {\it
generalized measure} on $X$, or that $(X,{\bf A}_0,\int)$ is a
{\it generalized measure space}.

We can do a large amount of measure theory abstractly in this context.
For example, if the generalized measure is positive:
\[          \int  \overline f f  \ge 0 ,\]
we can define $L^2(X,\int)$ to be the completion of ${\bf A}_0$ in the norm
\[                   \|f\|_2 = \left|\int \overline f f\right|^{1/2}  .\]
Suppose that $G$ is a group acting as homeomorphisms of $X$.  Then $G$
acts on $C_b(X)$ by
\[                 gf(x) = f(g^{-1}x) .\]
Suppose that this action preserves the generalized measure $\int$, that
is, $f \in {\bf A}_0$ implies
$gf  \in {\bf A}_0$, and $\int f = \int gf$.  Then the action of $G$ on
${\bf A}_0$ extends to a unitary action of $G$ on $L^2(X,\int)$.

On the other hand, if we wish, we can
translate the theory of generalized measures back into the theory
of measures on compact Hausdorff spaces, allowing us to use all the
standard tools of measure theory.  Suppose $(X,{\bf A}_0,\int)$ is a
generalized measure space.  Let ${\bf A}$ denote the completion of ${\bf
A}_0$ in the $L^\infty$ norm, or equivalently, the closure of ${\bf
A}_0$ in $C_b(X)$.  Then by the Gelfand-Naimark theorem there is a
compact Hausdorff space $\overline X$ such that ${\bf A} \iso
C(\overline X)$.  There is also a continuous map from $X$ to $\overline
X$ with dense range.  If ${\bf A}_0$ is
sufficient to separate points in $X$, that is, if $x \ne y$ implies there
exists $f \in {\bf A}_0$ with $f(x) \ne f(y)$, then the map from
$X$ to $\overline X$ is one-to-one.
Every function $f \in {\bf A}$ canonically defines a
function $\tilde f$ on $\overline X$.
On the other hand, since the functional $\int$ is bounded, it
 extends uniquely to a continuous linear
functional $\int \maps {\bf A} \to \C$.
Thus by the Riesz-Markov theorem there is a unique
measure $\mu$ on $\overline X$ such that
\[             \int f = \int_{\overline X} \,\, \tilde f d\mu  \]
for all $f \in {\bf A}$.

In short, generalized measure theory on $X$
can be translated into ordinary measure theory on a compact space
$\overline X$ containing certain ``limits'' of points of $X$.
This way of thinking also extends to the case when there is a group
action present.
If $G$ acts as homeomorphisms of $X$ and preserves the
generalized measure $\int$, there is a unique action of $G$ as
homeomorphisms of $\overline X$ such that the map from $X$ to $\overline
X$ is $G$-equivariant, and this action on $\overline X$ preserves the measure
$\mu$.
In the next sections we will apply this philosophy to the case in which
$X$ is the space $\A/\G$ of connections modulo gauge transformations for
some bundle over a manifold $M$, and seek generalized measures on
$\A/\G$ that are invariant under the action of $\Diff_0(M)$.

\section{The Analytic Holonomy C*-algebra}

Let $G$ be a compact Lie group, and $\rho$ a $k$-dimensional unitary
representation of $G$.  Let $M$ be a real-analytic $n$-manifold and $P \to M$
a principal $G$-bundle over $M$.   Define $\tau \maps G \to \C$
by
\[        \tau(g) = {1\over k}\tr(\rho(g)).\]
Given a
smooth connection $A$ on $P$ and a piecewise smooth loop $\gamma \maps
S^1 \to M$, let $T(\gamma,A)$ denote the trace of the holonomy of $A$
around the loop $\gamma$, computed using the trace $\tau$.
Let $\A$ denote the space of smooth connections on $P$ and
$\G$ the group of smooth gauge transformations of $P$.  The functions
$T(\gamma) = T(\gamma,\cdot)$, known as Wilson loops,
are $\G$-invariant bounded continuous functions on $\A$.  Alternatively,
they may be thought of as elements of $C_b(\A/\G)$.

In general, a holonomy algebra is a subalgebra of $C_b(\A/\G)$ generated
in some manner by the functions $T(\gamma)$ for some class of loops $\gamma$.
There are a variety of versions of holonomy algebra.
The original holonomy algebra due to Ashtekar and
Isham \cite{AI} was generated by the traces of holonomies around
all piecewise smooth loops, and their holonomy C*-algebra was the
completion of the holonomy algebra in the $L^\infty$ norm.   The
topology of a piecewise smooth loop can be extremely complicated, which
makes it difficult to construct diffeomorphism-invariant states on this
holonomy C*-algebra.
Here we will work with the holonomy algebra defined by Ashtekar and
Lewandowski \cite{AL}, which involves only piecewise analytic loops.
In Section 6 of this paper we will mention some other sorts of holonomy
algebra that involve ``regularized'' or ``smeared'' Wilson loops.  It
appears necessary to consider such holonomy algebras to treat the
Chern-Simons path integral as a generalized measure.

Henceforth we will assume $M$ to be a real-analytic manifold.
We say that $\gamma \maps S^1 \to M$ is {\it piecewise analytic} if
$\gamma$ is continuous and
we can write $S^1$ as a finite union of closed intervals $I_i$ such that
$\gamma|_{I_i}$ extends to a real-analytic function from an open interval
containing $I_i$.  From now on we
will use the word {\it loop} to mean a piecewise analytic loop.
Let $\H_0$ denote the algebra of functions on $\A$ generated by
the functions $T(\gamma)$ for all such loops,
and let $\H$ denote the closure of $\H$ in the norm
\[          \|f\|_\infty = \sup_{A \in \A} |f(A)|  .\]
Note that the pointwise complex conjugate of the function $T(\gamma)$ is
the function $T(\gamma^{-1})$, where $\gamma^{-1}$ is $\gamma$ with its
orientation reversed.  Thus $\H_0$ is closed under complex conjugation,
so $\H$ is a C*-subalgebra of the C*-algebra $C_b(\A/\G)$.

The algebra $\H$ is called the {\it
holonomy C*-algebra} of the associated bundle $P \times_\rho \C^n$.
By the general theory of the previous section, $\H$ is isomorphic to
$C(\overline{\A/\G})$ for some compact Hausdorff space $\overline{\A/\G}$, and
there is a continuous map from $\A/\G$ to $\overline{\A/\G}$ with
dense range.
When the the functions in $\H_0$
separate the points of $\A/\G$, the map from $\A/\G$ to
$\overline{\A/\G}$ is one-to-one.  This is true when $\rho$ is the
fundamental representation of $SU(n)$, and probably much more generally.
In these cases, points of $\overline{\A/\G}$
 may be regarded as gauge equivalence classes of
 ``distributional connections'' on $P$.   Ashtekar, Isham and Lewandowski have
given some explicit examples of such distributional connections, as well as a
very clean abstract description of all of them when $\rho$ is the
fundamental representation of $SU(2)$ \cite{AI,AL}.

The group $\Diff(M)$ of real-analytic diffeomorphisms
of $M$ acts as homeomorphisms of $\A/\G$, as
 $\ast$-automorphisms of $\H$ by
\[         gT(\gamma) = T(g\circ \gamma) ,\]
and dually on the space of
continuous linear functionals $\H^\ast$.
Let us write $\H^\ast_{inv}$ for the elements of
$\H^\ast$ that are invariant under the action of $\Diff_0(M)$, the identity
component of $\Diff(M)$.

In fact,
elements of $\H^\ast_{inv}$ are classified by multiloop invariants, where
by a {\it multiloop} we mean an $n$-tuple of loops.
Let us say that two multiloops $\gamma = \{\gamma_i\}$ and $\eta =
\{\eta_i\}$ are {\it ambient isotopic} if for some $g \in \Diff_0(M)$ we
have $\gamma_i = g\circ \eta_i$.

\begin{proposition} \label{prop1}\et  Suppose that $\mu \in
\H^\ast_{inv}$.  Then there is an ambient isotopy invariant
of multiloops $\L_\mu$ given by
\[             \L_{\mu} (\gamma) = \mu(T(\gamma_1) \cdots T(\gamma_n)) \]
for any multiloop $\gamma = \{\gamma_i\}$.  Moreover,
the map $\mu \mapsto \L_{\mu}$ is one-to-one.  \end{proposition}

Proof - That $\L_{\mu}$ is a multiloop invariant follows directly
from the definitions, and the map $\mu \mapsto \L_{\mu}$ is one-to-one
because $\H_0$ is dense in $\H$.  \qed

A fundamental unsolved problem in the theory of holonomy algebras is to
determine which multiloop invariants are of the form $\L_\mu$ for some
$\mu \in \H_{inv}^\ast$, for a fixed $G$-bundle $P$ and representation
$\rho$.   One obvious constraint is that $\L_\mu(\gamma) = \L_\mu(\eta)$
if $T(\gamma_i) = T(\eta_i)$ for all $i$.  This constraint has been
studied by Ashtekar and Isham \cite{AI}.  The fundamental representation
of $SU(2)$ is special in this regard, in that $T(\gamma) =
T(\gamma^{-1})$ for all loops $\gamma$.

Every multiloop invariant restricts to an ambient
isotopy invariant of oriented links in $M$, but this restriction map is
not one-to-one.  Thus multiloop invariants have more information than
link invariants; they depend not only on the topology of the mappings
$\gamma_i \maps S^1 \to M$ but also on the structure of the
singularities of these mappings.
In what follows we will construct elements of $\H^\ast_{inv}$
from data corresponding to
the various possible singularities, or ``vertex types,'' admitted by analytic
multiloops.   Our method generalizes that of Ashtekar and Lewandowksi, who
constructed a state $\mu \in \H^\ast_{inv}$ in the case when $\rho$ is the
fundamental representation of $G = SU(2)$.
To construct examples states
in  $\H^\ast_{inv}$, in the next section we develop
a characterization of all such states.

\section{Reduction to the Finite-Dimensional Case}

To tackle the problem of constructing elements of $\H^\ast_{inv}$, we can
adapt a technique used in functional integration on
infinite-dimensional vector spaces.  This is an algebraic
formulation of the idea of ``cylinder measures,'' which goes back to
Kolmogorov and has subsequently received many formulations
\cite{BSZ,Kolm}.  If $V$ is a real
vector space, let ${\bf A}_0$ be the algebra of
{\it tame} functions on $V$, that is, functions $f \maps V \to \C$ such
\[           f = \tilde f \circ j \]
where $j \maps V \to \R^d$ is a quotient map, that is, an onto linear
map, and $\tilde f \in C_b(\R^d)$.
In other
words, the tame functions are the bounded continuous functions depending
on only finitely many coordinates of $V$.  We can construct
generalized measures on $V$ as follows.  Suppose that for each
quotient map $j \maps V \to \R^d$ ($d$ arbitrary) there is given a measure
$\mu_j$ on $\R^d$.   We may attempt to define $\int \maps {\bf A}_0 \to
\C$ by setting
\[         \int f  =   \int_{\R^d} F d\mu_j \]
if $f = F \circ j$.
For $\int$ to be well-defined, the following
consistency condition is sufficient: if $f = F \circ j$ and also
$f = F' \circ j'$ for some other quotient map $j' \maps V \to R^{d'}$,
we must have
\[   \int_{\R^d} F d\mu_j =  \int_{\R^{d'}} F' d\mu_{j'} .\]
If in addition there is a constant $C > 0$  such that
\[        \left| \int_{R^d} F d\mu_j\right| \le C \|\tilde f\|_\infty\]
for all quotient maps $j$ and all $F \in C_b(\R^d)$,
$(V,\a_0,\int)$ is a generalized
measure space.  The advantage of this construction is that it reduces
integration on $V$ to a problem only involving finite-dimensional spaces.

Similarly, we will show how to construct elements of $\H^\ast_{inv}$
from a family of measures on spaces of
the form $G^d$ that satisfies consistency and uniform boundedness
conditions.  Conversely, we show that {\it all} elements of $\H^\ast_{inv}$
arise this way.  In the next section we construct some examples.

First, we define the notion of an embedded graph in $M$.
We define an
{\it embedded graph} in $M$ to be a nonempty set $\phi \subseteq M$ such
that there exist finitely many maps $\phi_i \maps [0,1] \to M$ with:

1)  $\phi = \bigcup_i \phi_i[0,1]$,

2)  each $\phi_i$ is one-to-one,

3) $\phi_i|_{(0,1)}$ is an embedding,

4) for all $i$ and $j$,
$\phi_i[0,1] \cap \phi_j[0,1] \subseteq \{\phi_i(0),\phi_j(0)\}$.

This implies that $\phi$ has the topology of a finite graph.
We call the maps $\phi_i$ {\it edges}.  Note that
for a given embedded graph $\phi$, which is just a set, there is not a unique
choice of edges $\phi_i$ satisfying 1) - 4).
We call a finite set of maps $\phi_i
\maps [0,1] \to M$ an {\it edge decomposition} of $\phi$ if 1) - 4) hold.

Given an embedded graph $\phi \subseteq M$ we define the subalgebra
$\a_0(\phi)$ of $\H_0$ to be the subalgebra generated by the elements
$T(\gamma)$ for all loops $\gamma \maps S^1 \to M$ with range lying in
$\phi$.  Let $\a(\phi) \subseteq \H$ denote the closure of $\a_0(\phi)$
in the $L^\infty$ norm.  Note that $\a_0(\phi)$ is a $\ast$-algebra and
$\a(\phi)$ is a C*-algebra.

If $\phi,\psi$ are embedded graphs with $\phi\subseteq\psi$, then
$\a(\phi) \subseteq \a(\psi)$.    We
will use the notation $\{\int_\phi\}$ to denote a family
of functionals $\int_\phi \in \a(\phi)^\ast$, one
for each embedded graph $\phi$.  We say such a family is {\it
consistent} if whenever $\phi\subseteq\psi$, then
$\int_\phi$ is the restriction of $\int_\psi$ to $\a(\phi)$.
We say the family is {\it uniformly
bounded} if for some $C > 0$,
$      \|\int_\phi \| \le C $ for all $\phi$.  We have:

\begin{theorem}\label{thm1}\et  Suppose that  $\int \in
\H^\ast$.  Given an embedded graph $\phi$, let $\int_\phi$
be the restriction of $\int$ to $\a(\phi)$.   Then  $\{\int_\phi\}$ is
 a consistent
and uniformly bounded family.
Conversely, given a consistent
and uniformly bounded family  $\{\int_\phi\}$
there exists a unique $\int \in
\H^\ast$ such that for all embedded graphs $\phi$, $\int_\phi$
is the restriction of $\int$ to $\a(\phi)$. \end{theorem}

Proof - Suppose that $\int \in
\H^\ast$ and let $\int_\phi$
be the restriction of $\int$ to $\a(\phi)$.
 Then the $\int_\phi$ are consistent and uniformly
bounded by the norm of $\int$.
Conversely suppose we are given a consistent and
uniformly bounded family $\{\int_\phi\}$.
In order to construct a functional $\int \in \H^\ast$ whose restriction
to each $\a(\phi)$ is $\int_\phi$, we use two lemmas whose proofs we
omit:

\begin{lemma} \label{lem1}\et If $\phi$, $\psi$ are embedded graphs, so
is $\phi\cup\psi$.  \end{lemma}

\begin{lemma} \label{lem2}\et If $\gamma \maps S^1 \to M$ is a loop,
the range of $\gamma$ is an embedded graph. \end{lemma}

Any element
$a \in \H_0$ is a finite linear combination of products of elements
$T(\gamma_i)$,
where $\{\gamma_i\}$ is a finite set of loops.  By Lemmas 1 and 2, there
is an embedded graph $\phi$ such that $T(\gamma_i) \in \a(\phi)$ for all
$i$.   Define
\[      \int f = \int_\phi f .\]
We need to check that $\int$ is well-defined, linear, and extends to a
continuous linear functional on $\H$.  Clearly the extension is unique
since $\H_0$ is dense in $\H$.

For well-definedness, suppose that $a \in  \a(\phi)$ and
$a \in \a(\psi)$ as well.   By Lemma 1, $\phi\cup\psi$ is an embedded
graph with $\phi, \psi\subseteq \phi\cup\psi$.  Thus by consistency,
\[       \int_\phi f = \int_{\phi\cup\psi} f = \int_\psi f .\]
For linearity, suppose $f,g \in \H_0$.  Then there are embedded graphs
$\phi,\psi$ such that $f \in \a(\phi)$ and $g \in \a(\psi)$.  Then
$f,g,f+g \in \a(\phi\cup\psi)$ and
\[        \int (f+ g) = \int_{\phi\cup\psi}(f+ g) =
  \int_{\phi\cup\psi}f + \int_{\phi\cup\psi} g = \int f + \int g.\]
Clearly $\int(\lambda f) = \lambda \int f$ for all $\lambda \in \C$.
Finally, to show that $\int$ extends  to a continuous linear functional on
$\H$ it suffices to note that for any $f \in \H_0$ we can choose
$\phi$ with $f \in \a(\phi)$, so
\[    |\int f| = |\int_\phi f| \le C \|f\|  .\]
\hskip 30em \qed

In physics one is especially interested in states on $\H$, so it is
useful to note that a consistent family of states on the subalgebras
$\a(\phi)$ determines a unique state on $\H$:

\begin{proposition} \label{prop2}\et If the family
 functionals $\{\int_\phi\}$ are consistent and
$\int_\phi$ is positive for every embedded graph $\phi$, then the family
$\{\int_\phi\}$ is uniformly bounded.       \end{proposition}

Proof - Note that the unit $1 \in \H$ is in $\a(\phi)$ for all $\phi$.
 Since $\int_\phi$ is positive, $\|\int_\phi\| = \int_\phi 1$.
So it suffices to show that for any two embedded graphs $\phi$ and $\psi$,
$\int_\phi 1 = \int_\psi 1$.  This follows from consistency:
\[     \int_\phi 1 = \int_{\phi\cup\psi} 1 = \int_\psi 1 .\]
\hskip 30em \qed

We now lay the groundwork for
constructing elements of $\H^\ast_{inv}$.  First note
that any diffeomorphism $g \maps M \to M$ with $g(\phi) \subset \psi$
for embedded graphs $\phi,\psi$ induces a $\ast$-homomorphism
$ g_\ast \maps \a_0(\phi) \to \a_0(\psi)$
such that
\[          g_\ast T(\gamma) = T(g\circ \gamma)  \]
for all loops $\gamma\maps S^1 \to M$ with range in $\phi$.  This
homomorphism is norm-preserving, so it is one-to-one, and extends to a
one-to-one $\ast$-homomorphism from $\a(\phi)$ to $\a(\psi)$, which we
also call $g_\ast$.  By duality we obtain a linear map
\[       g^\ast \maps \a(\psi)^\ast \to \a(\phi)^\ast  .\]
Given a family $\{\int_\phi\}$ of functionals on
the algebras $\a(\phi)$, we say it is {\it covariant} if for any $g \in
\Diff_0(M)$, given embedded graphs $\phi$ and $\psi$ with $g(\phi)
\subseteq \psi$, then
\[         g^\ast \int_\psi = \int_\phi  .\]
Note that a covariant family is automatically consistent.

\begin{theorem} \label{thm2} \et
 Suppose that  $\int  \in
\H_{inv}^\ast$.  Given an embedded graph $\phi$, let $\int_\phi$
be the restriction of $\int$ to $\a(\phi)$.   Then  $\{\int_\phi\}$ is
 a covariant
and uniformly bounded family.
Conversely, given a covariant
and uniformly bounded family  $\{\int_\phi\}$
there exists a unique  $\int \in
\H_{inv}^\ast$ such that for all embedded graphs $\phi$, $\int_\phi$
is the restriction of $\int$ to $\a(\phi)$. \end{theorem}

Proof - Using Theorem \ref{thm1}, the only substantial point to check is
that the element  $\int \in \H^\ast$ determined by a covariant and
uniformly bounded family is $\Diff_0(M)$-invariant.  For this it
suffices to check that $\int f = \int gf$ for all $f \in \H_0$ and $g
\in \Diff_0(M)$.
By Lemmas 1 and 2 we
can find an embedded graph $\phi$ such that $f \in \a(\phi)$.  The image
$g\phi$ of $\phi$ under $g$ is again an embedded graph and $ga \in
A(g\phi)$, so by covariance
\[     \int f = \int_\phi f = \int_{g\phi} gf = \int gf .\]
\hskip 30em \qed

Note that an embedded graph
$\phi$ has the topology of a finite graph with the
points $\phi_i(0), \phi_i(1)$ as vertices and the sets $\phi_i[0,1]$ as
edges.  Thus the
study of states on $\a(\phi)$ essentially amounts to doing lattice gauge
theory on a graph.    (For a review of applications
of Wilson loops to lattice gauge theory,
see Loll \cite{Loll}.)
Let $\pi_1 = \pi_1(\phi)$ denote the fundamental group of $\phi$, which
we define as the free product of the fundamental groups of the components
of $\phi$ if $\phi$ is not connected.
Then $\pi$
is a finitely generated free group, and for any edge decomposition of
$\phi$ we may find loops generating $\pi_1$ that are products of
the edges $\phi_i$ and their inverses.
Since $\phi$ is a graph, the holonomy of a connection $A \in \A$ around
any loop $\gamma \maps S^1 \to \phi$ depends only on its homotopy
class.  Thus we have a map
\[        p \maps \A \to \Hom(\pi_1,G)  ,\]
and as noted by Ashtekar and Lewandowski \cite{AL} this map
is onto when $G$ is connected, which we assume henceforth.
Note that $\Hom(\pi_1,G)$ is a manifold diffeomorphic to $G^d$, where
$d$ is the rank of $\pi_1$.  Moreover, any element of
$\a(\phi)$, regarded as a function on $\A$, is of the form $f \circ
p$ for some $f \in C(\Hom(\pi_1,G))$.  Since $p$ is onto, $f$ is unique.
Thus we may identify $\a(\phi)$ with a subalgebra of
$C(\Hom(\pi_1,G))$.
It follows that any measure on
$\Hom(\pi_1,G)$ determines an element $\int_\phi \in \a(\phi)^\ast$.

In the next section we sketch how to actually construct elements of
$\H_{inv}^\ast$ using Theorem 2.
The results from this section that we will need can be
expressed as follows.  We say that $\{\mu_\phi\}$ is a {\it covariant
family} of measures if for each embedded graph $\phi$, $\mu_\phi$
is a measure on $\Hom(\pi_1(\phi),G)$, and for all
$g \in \Diff_0(M)$ with $g \maps \phi \to \psi$ we have
\[        g^\ast \mu_\psi = \mu_\phi  ,\]
where $g^\ast$ is the
map from measures on $\Hom(\pi_1(\psi),G)$ to
measures on $\Hom(\pi_1(\phi),G)$ induced by $g$.  We have:

\begin{corollary} \label{cor1}   Suppose $G$ is connected, and
suppose that $\{\mu_\phi\}$ is a
covariant family of probability measures.
Then  there exists a unique state $\int \in \H^\ast_{inv}$ such that for all
embedded graphs $\phi$ and all $f \in \a(\phi)$,
\[            \int f = \int_{\Hom(\pi_1(\phi),G)} f d\mu  \]
where on the right side we identify $f$ with a function on
 $\Hom(\pi_1(\phi),G)$.  \end{corollary}

Proof - This follows immediately from Proposition \ref{prop2}, Theorem
\ref{thm2}, and the remarks above.  \qed

It is convenient to think of probability measures on
 $\Hom(\pi_1(\phi),G)$ in terms of $G$-valued random variables.
A  probability measure on $\Hom(\pi_1(\phi),G)$ is the same as an
function from elements $\gamma \in \pi_1(\phi)$
to $G$-valued random variables $g(\gamma)$ such that whenever
$\gamma,\eta$ are homotopy classes of loops in the same component of $\phi$,
\[           g(\gamma)g(\delta) = g(\gamma \delta) . \]
Since $\pi_1(\phi)$ is a free group with $d$ generators, we may also
think of a probability measure on $\Hom(\pi_1(\phi),G)$ as a $d$-tuple
of $G$-valued random variables, one for each generator.

\section{Constructing Diffeomorphism-invariant States}

The space of multiloops is a stratified space.  Embeddings, or
links, form an open dense subset of this space, while strata of
increasingly high codimension correspond to multiloops with ever more
complicated self-intersections and other singularities.  It seems likely
that a deeper understanding of the connection between knot theory and
gauge theory will require studying the whole space of multiloops, and
involve a blend of singularity theory and group representation theory.
There is quite an amount of work that points in this direction, even
though the full picture has not yet emerged.

On the one hand, Vassiliev's study of the space of loops \cite{Vas}
led him to formulate the notion of knot invariants of finite type.
Then Bar-Natan, Birman and Lin \cite{BN,Birman} found that such knot
invariants may be constructed from the perturbative
Chern-Simons theory, or alternatively via group
representations theory.
Further work showed that there really is a theory of graph invariants of
finite type \cite{Stanford}, and that finite type invariants
may be related to perturbative quantum gravity \cite{Baez}.

On the other hand,
in the loop representation of quantum gravity it appears that multiloop
invariants having trivial behaviour on multiloops with singularities
form only a small part of the space of states
\cite{BGP,Husain}.  Thus it is
important to devise systematic constructions of multiloop invariants,
and in particular, natural ways of extending known link
invariants to multiloops
with self-intersections.  This has been pursued for the Chern-Simons
link invariants by Br\"ugmann, Gambini, Pullin, Kauffman,
and others \cite{BGP2,Gambini,Kauffman}.

Here we outline a procedure to construct diffeomorphism-invariant generalized
measures on $\A/\G$ - or in other words, elements of $\H_{inv}^\ast$ -
using Corollary 1.
First we define an equivalence relation on points in embedded graphs.
Given embedded graphs $\phi,\psi$ and points $x \in \phi$, $y \in
\psi$, we say that $(x,\phi)$ and $(y,\psi)$ have the same {\it vertex
type} if
there is an element $g \in \Diff_0(M)$ such that $g(x) = y$ and there are
neighborhoods $U \ni x$, $V \ni y$ such that $g(\phi \cap U) = \psi\cap
V$.  We call an equivalence class of pairs $(x,\phi)$ a vertex
type, and write the vertex type containing $(x,\phi)$ as
$[x,\phi]$.  Some of the simpler vertex types (for $\dim M \ge 2$) are
the {\it half-arc}:
\begin{center}
\setlength{\unitlength}{0.00625in}%
\begin{picture}(10,45)(315,555)
\thicklines
\put(320,560){\circle*{10}}
\put(320,600){\line( 0,-1){ 40}}
\end{picture}
\end{center}
the {\it arc}:
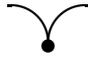
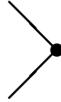
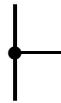
\begin{center}
\setlength{\unitlength}{0.00625in}%
\begin{picture}(10,80)(315,520)
\thicklines
\put(320,560){\circle*{10}}
\put(320,600){\line( 0,-1){ 40}}
\put(320,560){\line( 0,-1){ 40}}
\end{picture}
\end{center}
the {\it cusp}:
\begin{center}
\setlength{\unitlength}{0.0125in}%
\begin{picture}(40,27)(300,455)
\thicklines
\put(338,460){\oval( 36, 34)[tl]}
\put(303,460){\oval( 34, 34)[tr]}
\put(320,460){\circle*{5}}
\end{picture}
\end{center}
the {\it corner}:
\begin{center}
\setlength{\unitlength}{0.00625in}%
\begin{picture}(45,80)(280,420)
\thicklines
\put(320,460){\circle*{10}}
\put(280,500){\line( 1,-1){ 40}}
\put(320,460){\line(-1,-1){ 40}}
\end{picture}
\end{center}
the {\it T}:
\begin{center}
\setlength{\unitlength}{0.00625in}%
\begin{picture}(45,80)(295,400)
\thicklines
\put(300,440){\circle*{10}}
\put(300,480){\line( 0,-1){ 80}}
\put(300,440){\line( 1, 0){ 40}}
\end{picture}
\end{center}
and the {\it transverse double point}:
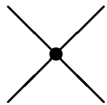
\begin{center}
\setlength{\unitlength}{0.0125in}%
\begin{picture}(40,40)(300,440)
\thicklines
\put(320,460){\circle*{5}}
\put(300,480){\line( 1,-1){ 40}}
\put(340,480){\line(-1,-1){ 40}}
\end{picture}
\end{center}
It is important to note that more complicated vertex types can come in
parametrized families.  For example, in
two dimensions there is a continuous one-parameter
family of vertex types that look roughly like:
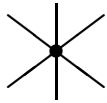
\begin{center}
\setlength{\unitlength}{0.0125in}%
\begin{picture}(40,40)(300,440)
\thicklines
\put(320,460){\circle*{5}}
\put(300,475){\line( 4,-3){ 40}}
\put(340,475){\line(-4,-3){ 40}}
\put(320,480){\line( 0,-1){ 40}}
\end{picture}
\end{center}

Given a edge decomposition $\phi_i$ of an embedded graph $\phi$, we
call the points $\phi_i(0),\phi_i(1)$
the {\it vertices}.  Note that the edges and vertices of $\phi$ depend
on the choice of edge decomposition.
For example, the following embedded graph $\phi$ has
an edge decomposition with 5 edges and 4 vertices:
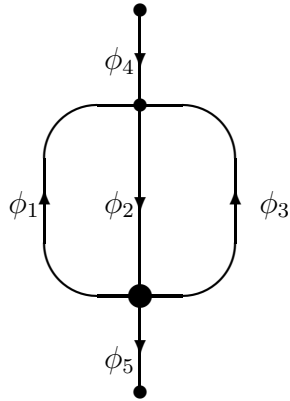
\begin{center}
\setlength{\unitlength}{0.0125in}%
\begin{picture}(105,160)(265,360)
\thicklines
\put(320,440){\oval( 80, 80)[br]}
\put(320,440){\oval( 80, 80)[tr]}
\put(320,440){\oval( 80, 80)[tl]}
\put(320,440){\oval( 80, 80)[bl]}
\put(320,400){\circle*{10}}
\put(320,480){\circle*{5}}
\put(320,520){\circle*{5}}
\put(320,360){\circle*{5}}
\put(280,435){\vector( 0, 1){ 10}}
\put(360,435){\vector( 0, 1){ 10}}
\put(320,480){\line( 0,-1){ 80}}
\put(320,445){\vector( 0,-1){ 10}}
\put(320,400){\makebox(0.4444,0.6667){\tenrm .}}
\put(320,520){\line( 0,-1){ 40}}
\put(320,400){\line( 0,-1){ 40}}
\put(320,505){\vector( 0,-1){ 10}}
\put(320,385){\vector( 0,-1){ 10}}
\put(265,435){\makebox(0,0)[lb]{\raisebox{0pt}[0pt][0pt]{$\phi_1$}}}
\put(305,435){\makebox(0,0)[lb]{\raisebox{0pt}[0pt][0pt]{$\phi_2$}}}
\put(370,435){\makebox(0,0)[lb]{\raisebox{0pt}[0pt][0pt]{$\phi_3$}}}
\put(305,495){\makebox(0,0)[lb]{\raisebox{0pt}[0pt][0pt]{$\phi_4$}}}
\put(305,370){\makebox(0,0)[lb]{\raisebox{0pt}[0pt][0pt]{$\phi_5$}}}
\end{picture}
\end{center}
We could, however, insert extra vertices, hence extra edges.

To construct $\Diff_0(M)$-invariant states on the holonomy algebra $\H$,
it suffices by Corollary 1 to
construct a covariant family
$\{\mu_\phi\}$, where $\mu_\phi$ is a
probability measures on $\Hom(\pi_1(\phi),G)$.
Defining $\pi_1(\phi)$ requires a choice of basepoint for each component
of $\phi$.  We may always choose these basepoints to be vertices; in the
example above we have arbitrarily chosen the vertex $\phi_2(1)$
as a basepoint for $\phi$.

We define the {\it valence} of a vertex type as in graph
theory, so that associated to each $n$-valent vertex type $v$ there is a
set $E(v)$ of $n$ (equivalence classes of) edges.
We will construct
 $\Diff_0(M)$-invariant states on the holonomy algebra $\H$ by a
procedure that involves fixing for each vertex type $v$ a
probability measure $m_v$ on $G^{E(v)}$, the set of maps from
$E(v)$ to the group $G$.   We may think of $m_v$ as a collection of
$G$-valued random variables, one for each edge in $E(v)$.

Suppose now that $x$ is any
vertex of the embedded graph $\phi$ (relative
to fixed edge decomposition $\{\phi_i\}$).  Let $v$ be the vertex type of
$x$, that is, let $v = [x,\phi]$.
We say an edge $\phi_i$ is {\it incident} to
$x$ if $x = \phi_i(0)$ or $x = \phi_i(1)$.
There is an obvious
one-to-one correspondence between the edges incident to $x$ and the set
$E(v)$.  This allows us to associate to each edge incident to $x$ a
$G$-valued random variable, such that the random variables
for all the edges incident to $x$
are distributed according to the probability measure $m_v$ on
$G^{E(v)}$.  We draw these random variables as dots near $x$ on the
edges incident to $x$, as follows:
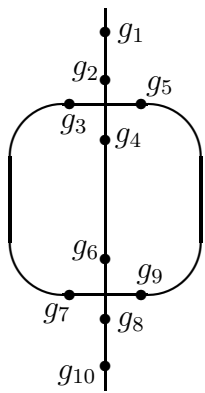
\begin{center}
\setlength{\unitlength}{0.0125in}%
\begin{picture}(136,160)(225,360)
\thicklines
\put(320,510){\circle*{4}}
\put(320,490){\circle*{4}}
\put(335,480){\circle*{4}}
\put(305,480){\circle*{4}}
\put(320,415){\circle*{4}}
\put(320,390){\circle*{4}}
\put(305,400){\circle*{4}}
\put(335,400){\circle*{4}}
\put(320,370){\circle*{4}}
\put(320,465){\circle*{4}}
\put(320,440){\oval( 80, 80)[br]}
\put(320,440){\oval( 80, 80)[tr]}
\put(320,440){\oval( 80, 80)[tl]}
\put(320,440){\oval( 80, 80)[bl]}
\put(320,480){\line( 0,-1){ 80}}
\put(320,480){\line( 0,-1){ 80}}
\put(320,400){\makebox(0.4444,0.6667){\tenrm .}}
\put(320,520){\line( 0,-1){ 40}}
\put(320,400){\line( 0,-1){ 40}}
\put(325,508){\makebox(0,0)[lb]{\raisebox{0pt}[0pt][0pt]{$g_1$}}}
\put(300,365){\makebox(0,0)[lb]{\raisebox{0pt}[0pt][0pt]{$g_{10}$}}}
\put(301,470){\makebox(0,0)[lb]{\raisebox{0pt}[0pt][0pt]{$g_3$}}}
\put(294,391){\makebox(0,0)[lb]{\raisebox{0pt}[0pt][0pt]{$g_7$}}}
\put(337,486){\makebox(0,0)[lb]{\raisebox{0pt}[0pt][0pt]{$g_5$}}}
\put(306,492){\makebox(0,0)[lb]{\raisebox{0pt}[0pt][0pt]{$g_2$}}}
\put(324,464){\makebox(0,0)[lb]{\raisebox{0pt}[0pt][0pt]{$g_4$}}}
\put(306,417){\makebox(0,0)[lb]{\raisebox{0pt}[0pt][0pt]{$g_6$}}}
\put(325,387){\makebox(0,0)[lb]{\raisebox{0pt}[0pt][0pt]{$g_8$}}}
\put(333,407){\makebox(0,0)[lb]{\raisebox{0pt}[0pt][0pt]{$g_9$}}}
\end{picture}
\end{center}
We do this for all the vertices, and require that
the random
variables associated to different vertices are independent.

For example,
the vertices $\phi_2(0)$ and $\phi_2(1)$ above have
the vertex type of the crossing, so we attach four $G$-valued
random variables to each, with distribution equal to the
the probability measure $m_+$ on $G^4$ associated to the crossing.
The vertices $\phi_4(0)$ and $\phi_5(1)$ have the vertex type of the
half-arc, so we attach one $G$-valued random variable to each,
with distribution equal to the probability measure $m_\halfarc$
associated to the half-arc.  Since the random variables near different
vertices are independent, the ten variables $g_j$
have distribution $m_\halfarc\tensor
m_+ \tensor m_+\tensor m_\halfarc$ on $G^{10}$.

This construction
allows us to associate a $G$-valued random variable to
each loop in $\phi$ that is a product of the edges $\phi_i$ and their
inverses.
Going around such a loop we write down a factor of $g_j$ for each dot we
come to while exiting a vertex, and a factor of $g_j^{-1}$ for each dot
we come to while entering a vertex.  For
example, to the loop $\phi_1\phi_2$ we associate the product
$g_7g_3^{-1}g_4g_6^{-1}$, while to the loop $\phi_5\phi_5^{-1}$ we associate
the product $g_8g_{10}g_{10}^{-1}g_8^{-1} = 1$.  Note that homotopic
loops automatically receive the same random variable.   Moreover,
this procedure associates to each homotopy class $\gamma \in
\pi_1(\phi)$ a random variable $g(\gamma)$ in such a way that
$g(\gamma)g(\eta) = g(\gamma\eta)$ whenever the product of homotopy
classes $\gamma,\eta$ is defined.   Thus, by the remark at the end of
the previous section, we have constructed a probability
measure $\mu_\phi$ on $\Hom(\pi_1(\phi),G)$.

It remains to state conditions on the measures $m_v$ such that the
measure $\mu_\phi$ is independent of a choice of edge decomposition
for the embedded graph $\phi$, and such that the family $\{\mu_\phi\}$ is
covariant in the sense of Corollary \ref{cor1}.  We will need two
conditions.

We define an {\it inclusion} of one vertex type in
another, written $i \maps [x,\phi] \to [y,\psi]$, to be an equivalence
class of diffeomorphisms $g \in \Diff_0(M)$ such that $g(\phi) \subseteq
\psi$ and $g(x) = y$, where we say two diffeomorphisms are equivalent if
they give rise to the same map from $E[x,\phi]$ to $E[y,\psi]$.
For example, there
are two inclusions of the half-arc in the arc:
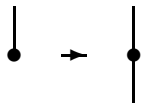
\begin{center}
\setlength{\unitlength}{0.00625in}%
\begin{picture}(110,80)(215,420)
\thicklines
\put(320,460){\circle*{10}}
\put(220,460){\circle*{10}}
\put(320,500){\line( 0,-1){ 80}}
\put(220,500){\line( 0,-1){ 40}}
\put(260,460){\vector( 1, 0){ 20}}
\end{picture}
\end{center}
and
\begin{center}
\setlength{\unitlength}{0.00625in}%
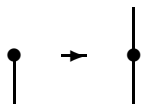
\begin{picture}(110,80)(215,420)
\thicklines
\put(320,460){\circle*{10}}
\put(220,460){\circle*{10}}
\put(320,500){\line( 0,-1){ 80}}
\put(260,460){\vector( 1, 0){ 20}}
\put(220,460){\line( 0,-1){ 40}}
\end{picture}
\end{center}
but no inclusions of the half-arc in the cusp.
Similarly, there are four inclusions of the arc in the transverse double
point, but no inclusions of the arc in the corner.
Note that given an
inclusion $i\maps v \to w$,  there is a natural inclusion of sets $E(v)
\hookrightarrow E(w)$.  This
in turn gives rise to a natural surjection $G^{E(w)} \to
G^{E(v)}$, and we can push forward a measure $\mu_w$ on $G^{E(w)}$ to a
measure which we call $i^\ast\mu_w$ on $G^{E(v)}$.
Our first condition is that
given any inclusion $i \maps v \to w$,
\be i^\ast m_w = m_v .       \label{cond1}  \ee

A given embedded graph typically has many edge decompositions.  However,
given any two edge decompositions of $\phi$ with vertices $\{x_i\}$, $\{y_i\}$
respectively, there is an edge decomposition with vertices $\{x_i\} \cup
\{y_i\}$.  In fact, one can go between any two edge decompositions by a
series of local moves in which one replaces
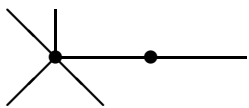
\begin{center}
\setlength{\unitlength}{0.0125in}%
\begin{picture}(100,40)(260,440)
\thicklines
\put(280,460){\circle*{5}}
\put(320,460){\circle*{5}}
\put(280,460){\line( 1, 0){ 40}}
\put(320,460){\line( 1, 0){ 40}}
\put(280,460){\line(-1, 1){ 20}}
\put(280,460){\line(-1,-1){ 20}}
\put(280,460){\line( 1,-1){ 20}}
\put(280,460){\line( 0, 1){ 20}}
\end{picture}
\end{center}
by
\begin{center}
\setlength{\unitlength}{0.0125in}%
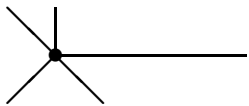
\begin{picture}(100,40)(260,440)
\thicklines
\put(280,460){\circle*{5}}
\put(280,460){\line( 1, 0){ 40}}
\put(320,460){\line( 1, 0){ 40}}
\put(280,460){\line(-1, 1){ 20}}
\put(280,460){\line(-1,-1){ 20}}
\put(280,460){\line( 1,-1){ 20}}
\put(280,460){\line( 0, 1){ 20}}
\end{picture}
\end{center}
or vice versa.  Here the vertex on the left stands for one of any type
and the vertex on the right has the type of an arc.
Our second condition is therefore as follows.
Suppose the random variables $(g,h_1,h_2)$ have distribution
 $m_v \tensor m_\arc$ on $G^3$, where $m_v$
is the measure on $G$ associated to any $1$-valent vertex type and $m_\arc$
is the measure on $G^2$ associated to the arc.
Let $p \maps G \times G^2 \to G$ be given by
\[        p(g,h_1,h_2) =  gh_1^{-1}h_2 .\]
Then we require
\be           p_\ast (m_v \tensor m_\arc) = m_v \label{cond2}  \ee
Pictorially, this says that
\[
    \setlength{\unitlength}{0.0125in}%
\begin{picture}(100,40)(260,440)
\thicklines
\put(270,460){\circle*{5}}
\put(310,460){\circle*{5}}
\put(270,460){\line( 1, 0){ 40}}
\put(310,460){\line( 1, 0){ 40}}
\end{picture}
\raise 3ex\hbox{\quad =\quad}
\setlength{\unitlength}{0.0125in}%
\begin{picture}(100,40)(260,440)
\thicklines
\put(280,460){\circle*{5}}
\put(280,460){\line( 1, 0){ 40}}
\put(320,460){\line( 1, 0){ 40}}
\end{picture}
\]   From conditions (\ref{cond1}) and (\ref{cond2}), it follows that
for any $n$-valent vertex type $v$, labelling the edges of $v$
arbitrarily with integers $\{1,\dots,n\}$, so that $m_v$ becomes a
probability measure on $G^n$, and defining
\[     p(g_1, \dots, g_n,h_1,h_2) = (g_1, \dots, g_{n-1},
g_nh_1^{-1}h_2),\]
we have
\[   p_\ast (m_v \tensor m_\arc) = m_v   .\]
Pictorially, this says that
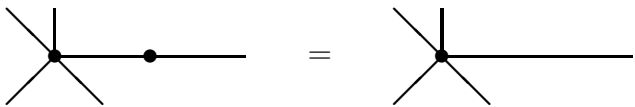
\[
\setlength{\unitlength}{0.0125in}%
\begin{picture}(100,40)(260,440)
\thicklines
\put(280,460){\circle*{5}}
\put(320,460){\circle*{5}}
\put(280,460){\line( 1, 0){ 40}}
\put(320,460){\line( 1, 0){ 40}}
\put(280,460){\line(-1, 1){ 20}}
\put(280,460){\line(-1,-1){ 20}}
\put(280,460){\line( 1,-1){ 20}}
\put(280,460){\line( 0, 1){ 20}}
\end{picture}
\raise 3ex\hbox{\qquad =\qquad}
\setlength{\unitlength}{0.0125in}%
\begin{picture}(100,40)(260,440)
\thicklines
\put(280,460){\circle*{5}}
\put(280,460){\line( 1, 0){ 40}}
\put(320,460){\line( 1, 0){ 40}}
\put(280,460){\line(-1, 1){ 20}}
\put(280,460){\line(-1,-1){ 20}}
\put(280,460){\line( 1,-1){ 20}}
\put(280,460){\line( 0, 1){ 20}}
\end{picture}
\]
This guarantees that the measures $\mu_\phi$ are independent of edge
decomposition.

In fact, condition (\ref{cond1}) also implies the
measures $\{\mu_\phi\}$ are a covariant family.  Suppose $\phi,\psi$ are
embedded graphs and $g \in \Diff_0(M)$ has $g(\phi) \subseteq \psi$.
Then we can find an edge decomposition of $\psi$ such that a subset of
the edges $\{\psi_i\}$ give an edge decomposition of $g\phi$.
Then for each vertex $x$ of $\phi$, $g(x)$ is a vertex of $\psi$, and
$g$ determines an inclusion $i \maps [x,\phi] \to [g(x),\psi]$.
Using these facts one easily sees from (\ref{cond1}) that
$g^\ast \mu_\psi = \mu_\phi$.

The simplest example of a family of probability measures $\{m_v\}$ meeting
conditions  (\ref{cond1}) and (\ref{cond2}) is that assigning to each vertex
type $v$ the Dirac delta at the identity of $G^{E(v)}$.  This gives the
{\it flat state} in $\H^\ast_{inv}$.
When the principal bundle $P$ admits a connection $A_0$ such
that the holonomy around every loop is trivial, i.e., when $P$ is
trivial, the flat state may be identified with Dirac delta
measure at $[A_0] \in \A/\G$.    Curiously, the flat state is
well-defined for any principal bundle $P$, even if $P$ admits no flat
connections.  The real reason for this is that the restriction of $P$ to any
embedded graph $\phi$ is trivial, since we are assuming that $G$ is
connected.   This permits a remarkable degree of ``bundle-independence''
for elements of $\H_{inv}^\ast$, a phenomenon noted by Ashtekar and
Lewandowski \cite{AL} that deserves further study.

A more interesting example of a family $\{m_v\}$ meeting
conditions  (\ref{cond1}) and (\ref{cond2}) is that assigning
to each vertex type $v$
the normalized Haar measure on the Lie group $G^{E(v)}$.
This gives the {\it Ashtekar-Lewandowski} state,
constructed by these authors in
the case where $G = SU(2)$ and $\rho$ is the fundamental representation
\cite{AL}.    It is easy to check that this state is
is faithful, that is, if $f \in \H$ has $f \ge 0$ and $\int_\phi f = 0$, then
$f = 0$.

A complete analysis of the solutions of
(\ref{cond1}) and (\ref{cond2}) would require a deep understanding of the
interplay of singularity theory for curves and harmonic analysis on $G$
involved in these equations.  We do not have such an understanding
yet - indeed, one might worry that the solutions we have mentioned are
the only ones!  To allay such fears, we present a few new solutions, but
leave as an open problem a thorough analysis of the conditions
(\ref{cond1}) and (\ref{cond2}).

It is easiest to describe solutions to (\ref{cond1}) and (\ref{cond2})
in terms of the $n$-tuple of $G$-valued random variables
assigned by $m_v$ to each $n$-valent vertex type $v$.
Given two edges $e,f \in E(v)$, let us say that $e$ and $f$ {\it form an
arc} if there is an inclusion $i \maps \higharc \to v$ of the arc in $v$
such that image of $E(\higharc)$ under the corresponding inclusion
$E(\higharc) \hookrightarrow E(v)$ is $\{e,f\}$.
For example, for the vertex below, a combination of the cusp and the T,
only edges $e_3$ and $e_5$ form an arc.
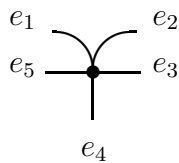
\begin{center}
\setlength{\unitlength}{0.0125in}%
\begin{picture}(55,64)(290,425)
\thicklines
\put(338,460){\oval( 36, 34)[tl]}
\put(303,460){\oval( 34, 34)[tr]}
\put(320,460){\circle*{5}}
\put(320,460){\line( 1, 0){ 20}}
\put(320,460){\line(-1, 0){ 20}}
\put(320,460){\line( 0,-1){ 20}}
\put(285,480){\makebox(0,0)[lb]{\raisebox{0pt}[0pt][0pt]{$e_1$}}}
\put(345,480){\makebox(0,0)[lb]{\raisebox{0pt}[0pt][0pt]{$e_2$}}}
\put(345,460){\makebox(0,0)[lb]{\raisebox{0pt}[0pt][0pt]{$e_3$}}}
\put(315,425){\makebox(0,0)[lb]{\raisebox{0pt}[0pt][0pt]{$e_4$}}}
\put(285,460){\makebox(0,0)[lb]{\raisebox{0pt}[0pt][0pt]{$e_5$}}}
\end{picture}
\end{center}
Now fix a probability measure $m$ on $G$.  To $v$ assign $n$ $G$-valued
random variables $g_e$, one for each edge $e \in E(v)$, that are all
distributed according to the measure $m$, and such that $g_e = g_f$ if
the edges $e$ and $f$ form an arc, while $g_e$ and $g_f$ are
independent otherwise.    Conditions (\ref{cond1}) and (\ref{cond2}) are
easy to check.  We thus obtain $\Diff_0(M)$-invariant states on $\H$.
Note that if
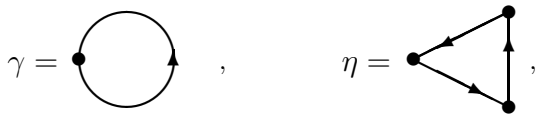
\begin{center}
\setlength{\unitlength}{0.0125in}%
\begin{picture}(215,50)(250,435)
\thicklines
\put(280,460){\circle*{5}}
\put(300,460){\circle{42}}
\put(420,460){\circle*{5}}
\put(460,440){\circle*{5}}
\put(460,480){\circle*{5}}
\put(420,460){\line( 2, 1){ 40}}
\put(460,480){\line(-2,-1){ 40}}
\put(420,460){\line( 2,-1){ 40}}
\put(460,480){\line( 0,-1){ 40}}
\put(430,455){\line( 0, 1){  0}}
\put(430,455){\vector( 2,-1){ 20}}
\put(460,450){\vector( 0, 1){ 20}}
\put(450,475){\vector(-2,-1){ 20}}
\put(340,455){\makebox(0.4444,0.6667){\tenrm ,}}
\put(470,455){\makebox(0.4444,0.6667){\tenrm ,}}
\put(319.5,459){\vector( 0, 1){  5}}
\put(250,455){\makebox(0,0)[lb]{\raisebox{0pt}[0pt][0pt]{$\gamma = $}}}
\put(390,455){\makebox(0,0)[lb]{\raisebox{0pt}[0pt][0pt]{$\eta = $}}}
\end{picture}
\end{center}
then $\int  T(\gamma) = 1$ but generally $\int T(\eta) \ne 1$ in these
states.  In the flat state, $\int T(\gamma) = 1$ for any loop $\gamma$.
In the Ashtekar-Lewandowski state for $\rho$ the fundamental
representation of $G = SU(2)$, $\int  T(\gamma) = \int T(\eta) = 0$.

\section{Generalizations and Conclusions}

When $M$ is 3-dimensional,
the construction of states in the previous section may be
regarded as the ``trivially braided'' case of a more general
construction.  If $M = \R^3$ or $S^3$, for example, we may assign $G$-valued
random variables not only to vertices but also to crossings in a planar
diagram of the embedded graph:
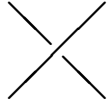
\begin{center}
\setlength{\unitlength}{0.00625in}%
\begin{picture}(80,80)(320,420)
\thicklines
\put(400,500){\line(-1,-1){ 80}}
\put(320,500){\line( 1,-1){ 35}}
\put(365,455){\line( 1,-1){ 35}}
\end{picture}
\end{center}
The conditions for the measure $\mu_\phi$ associated to a given embedded
graph $\phi$ to be independent of the choice of diagram will involve the
Yang-Baxter equation and relations allowing us to move vertices past
crossings, e.g.:
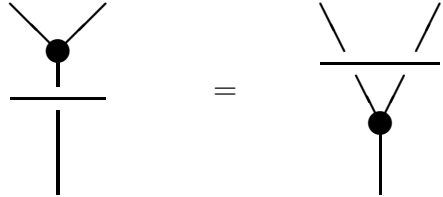
\begin{center}
\setlength{\unitlength}{0.0125in}%
\begin{picture}(180,80)(320,440)
\thicklines
\put(340,500){\circle*{10}}
\put(475,470){\circle*{10}}
\put(320,520){\line( 1,-1){ 20}}
\put(340,500){\line( 1, 1){ 20}}
\put(340,500){\line( 0,-1){ 15}}
\put(320,480){\line( 1, 0){ 40}}
\put(340,475){\line( 0,-1){ 35}}
\put(450,520){\line( 1,-2){ 10}}
\put(465,490){\line( 1,-2){ 10}}
\put(475,470){\line( 1, 2){ 10}}
\put(490,500){\line( 1, 2){ 10}}
\put(450,495){\line( 1, 0){ 50}}
\put(475,470){\line( 0,-1){ 30}}
\put(405,480){\makebox(0,0)[lb]{\raisebox{0pt}[0pt][0pt]{$=$}}}
\end{picture}
\end{center}

These conditions are, in fact, very similar to the Reidemeister-type moves
for rigid-vertex isotopy of graphs in the sense of Kauffman
\cite{Kauff2}, as well as the moves Reshetikhin and Turaev discuss in
their work on ribbon graphs \cite{RT}.  Reshetikhin and Turaev
systematically construct
invariants of ribbon graphs using braided tensor categories, of
which categories of representations of quantum groups are the main
example.
In particular, the quantum group link invariants are
precisely those one would hope to obtain from the Chern-Simons path
integral.  The quantum group link invariants depend on a parameter $q$,
which is related to the
integer $k$ appearing in the Chern-Simons action
by
\[                q = e^{2\pi i/(k + c_2(G)/2)}  \]
where $c_2(G)$ is the value of the quadratic Casimir operator for $G$ in
the adjoint representation, normalized so that it equals $2n$ for
$SU(n)$.   Ideally, one would hope to be able to construct the
Chern-Simons ``measure'' as a generalized measure on $\A/\G$ using the
sort of construction of the previous section, but with a nontrivial
braiding.   In particular, one would
hope that, at least for certain values of the quantization parameter
$q$, the universal $R$-matrix for the quantum group associated to a
semisimple Lie group $G$ could be expressed as a measure on
$G \times G$.  This appears to be an open question.

The main difficulty in trying to construct the Chern-Simons ``measure''
as an element of $\H^\ast_{inv}$ is that the link
invariants associated to the Chern-Simons path integral are
framing-dependent, while elements of $\H^\ast_{inv}$
determine framing-independent link invariants.   In certain cases this
problem can be sidestepped, since the framing dependence typically
enters via an exponential of the writhe, or self-linking number, of the
framed link.   Namely, if one works with
a Lie group of the form $G \times U(1)$, one can arrange that the
$U(1)$ factor provides an exponential of the writhe that cancels that
coming from $G$ \cite{BN}.
For example, Chern-Simons theory in the fundamental
representation of $G = SU(2)$
gives the Kauffman bracket, an invariant of framed unoriented links, but
Chern-Simons theory with $SU(2) \times U(1)$ can be used to obtain the
Jones polynomial, an framing-independent invariant of oriented links
that differs from the Kauffman bracket by a factor of an exponential of
the writhe.

For more
general Chern-Simons path integrals in which we cannot arrange a
cancellation of factors involving the writhe, we will need to replace the
holonomy algebra $\H$ by one for which diffeomorphism-invariant
continuous linear functionals yield invariants of framed links.
This is was the motivation for earlier work
\cite{Baez2} in which we described a modified holonomy algebra generated
by regularized Wilson loops, or {\it tubes}: functions on $\A/\G$ of
the form
\[               \int_{D^{n-1}}       T(\gamma^x,A)\, \om(x)    \]
where $\om$ is a smooth $(n-1)$-form compactly supported in the interior
of $D^{n-1}$, $\gamma \maps S^1 \times D^{n-1} \to M$ is a smoothly
embedded torus in the smooth manifold $M$, and for each $x \in D^{n-1}$
the loop $\gamma^x$ is given by
\[      \gamma^x(t) = \gamma(t,x)  .\]
The completion of this algebra in the $L^\infty$ norm is called the {\it
tube C*-algebra}.
There is a linear map from  $\Diff_0(M)$-invariant continuous linear
functionals on the tube C*-algebra to ambient isotopy
invariants of framed links.  What is more, this map is
one-to-one.  Thus the regularization involved in working with the tube
algebra has two good effects: it reduces the amount of information
contained in a $\Diff_0(M)$-invariant continuous linear functional from
a multiloop invariant to a link invariant, and it introduces the
possibility of framing-dependence.

Unfortunately, it appears difficult to construct $\Diff_0(M)$-invariant
continuous linear functionals on the tube algebra by methods analogous
to those of the present paper, essentially because tubes are too
``thick'' for the method of embedded graphs to apply.
A promising compromise currently under investigation is the ``strip algebra''
based on analytically embedded annuli $\gamma \maps S^1 \times [0,1] \to
M$.

In addition to these directions for further investigation,
it is tempting to try to construct the Chern-Simons measure not just on
$S^3$, but on general compact
3-manifolds using the machinery of modular tensor categories
\cite{Crane,RT}.   Interestingly, and not at all coincidentally, the
category of representations of a quantum group gives rise to a modular
tensor category precisely when
\[                q = e^{\pm 2\pi i/( k + c_2(G)/2)}  \]
with $k$ a nonnegative integer.   It would also be
interesting, and comparatively straightforward, to extend the theory
developed in this paper to the case of manifolds with boundary,
generalizing from Wilson loops to include also Wilson lines with
endpoints at the boundary, in order to make contact with the theory of
tangles and braid group representations \cite{BaezTang}.

To conclude, it should be clear that holonomy algebras offer a promising
route to doing diffeomorphism-invariant gauge theory in a rigorous way.
There is much to be done to explore the connections
between topology, singularity theory, representation theory and
category theory that arise in the study of diffeomorphism-invariant
states on holonomy C*-algebras.

\vskip 1em
{\it Acknowledgements.}  I would like to thank Louis Crane and David
Yetter for inviting me to speak at this conference, and Abhay Ashtekar,
Jerzy Lewandowski, and Lee Smolin for many useful discussions concerning
holonomy algebras and the loop representation of quantum gravity.
In particular, I would like to thank Ashtekar and Lewandowski for
showing me a draft of their paper on the holonomy C*-algebra.
Subsequent to writing this paper, I found that they had independently
developed similar ideas on using graphs to generalize the notion of
cylinder measures to the context of gauge theory.

\vfill
\end{document}